# On the microplasticity and dynamic strain aging in an FeAlCrMo complex-concentrated alloy

Tomáš Tayari[1], Michal Knapek[1,*], Eliška Jača[1], Peter Minárik[1], Josef Pešička[1]

[1] Charles University, Faculty of Mathematics and Physics, Department of Physics of Materials, Ke Karlovu 5, 121 16 Prague, Czech Republic

* Corresponding author: michal.knapek@matfyz.cuni.cz

**Abstract**

We show by acoustic emission analysis that FeAlCrMo complex-concentrated alloy (CCA) exhibits signatures of self-organization of deformation processes during both microplasticity and serrated flow (dynamic strain aging). Due to complex microstructures of CCAs and scarcity of literature data, these novel alloys are a prominent subject of future research efforts.

**Introduction**

Complex-concentrated alloys, as contemporarily rejuvenated in the early 2000s [1,2], have obtained significant attention from materials scientists. These alloys feature a suite of attractive properties, such as outstanding yield and fracture stress, very good wear resistance, and excellent oxidation and corrosion resistance, all at even relatively high temperatures [3–6]. On the other hand, it might be difficult to attain a combination of these properties in individual CCAs. Particularly, it has been challenging to effectively predict and design CCAs with the desired performance due to the microstructural complexity resulting from the presence of at least three elements at high concentrations.

It might be assumed that deformation dynamics in these (either single- or multiphase) materials can be rather complex at the microstructural level. Put simply, the dislocation motion might be influenced by strong lattice distortions related to different atomic radii, short-range ordering, or the presence of phase interfaces [7]. Nonetheless, the deformation curves of CCAs usually qualitatively resemble those of conventional diluted alloys [4]. Recently, it was also reported that macroscopic plastic instabilities (serrated plastic flow) can occur in certain CCAs [7–9], like in some diluted alloy systems, resembling the known serrated flow dynamics [10,11].

In our recent work [12], we documented the occurrence of macroscopic serrated flow (also known as the Portevin Le-Châtelier (PLC) effect) in the as-cast equiatomic FeAlCrMo CCA and explained it in terms of dynamic strain aging (DSA). DSA is a microscale effect caused by repeated pinning of dislocation cores by the diffusing solutes, and appears only within a specific range of strain rates and temperatures [13]. We confirmed the presence of (i) negative strain rate sensitivity, (ii) strain rate and temperature dependence of the serration type, and (iii) correlation of stress drops with acoustic emission (AE) signals, i.e the well-known signatures of DSA and the resulting PLC effect in some common diluted alloys.

In this letter, we report details on intriguing features evidenced by the analysis of AE signals of the equiatomic FeAlCrMo CCA recorded during compression at elevated temperatures. Namely, (i) we identify and examine microplasticity, i.e. the local plastic events taking place early in the (quasi) elastic stage of loading, and (ii) we inspect statistical features of avalanche-like characteristics of dislocation motion, manifesting themselves as power-law distributions of the magnitude of elementary plastic events. Along these lines, we assess the validity of the universality of these relatively newly explored and essential concepts across different material classes and make an extension to microstructurally complex and less investigated CCAs.

**Methods**

Ingots of equiatomic FeAlCrMo alloy were prepared by repeated arc melting under Ar atmosphere. The chemical composition of the as-cast alloy was determined by atomic absorption spectrometry to be (in at.%): 25.0 ± 0.3 Fe, 26.1 ± 0.3 Al, 24.3 ± 0.3 Cr, and 24.6 ± 0.3 Mo. The microstructure of the studied samples was investigated by scanning electron microscope (SEM) Zeiss Auriga Compact (ZEISS, Oberkochen, Germany) equipped with an electron backscatter diffraction (EBSD) detector (EDAX, Mahwah, NJ, USA). The acceleration voltage and working distance were 30 kV and 12 mm, respectively, for both SEM and EBSD measurements. The EBSD maps were measured with a step size of 1 μm. For the SEM/EBSD analysis, the samples were ground using abrasive SiC papers up to grade 2000 and subsequently vibratory-polished down to 50 nm alumina suspension. Further details on the preparation procedure and microstructural properties were reported elsewhere [14].

Compression tests were performed on the samples with dimensions of 7.5 × 5 × 5 mm$^3$ using the Instron 5882 machine at 300 °C and 400 °C with an initial strain rate of $10^{-4}$ s$^{-1}$. Several tests were performed to verify the repeatability of the results, and only representative data are shown here. The compression tests were complemented by in-situ AE monitoring using the AMSY-6 AE system (Vallen Systeme GmbH, Wolfratshausen, Germany), the broadband high-temperature AE sensor S9215 (Physical Acoustics Corporation (PAC), NJ, USA), and the PAC preamplifier, model 2/4/6, with a gain set to 60 dB. The AE streaming data were recorded with a sampling rate of 2 MHz and a band-pass filter of 0.1-1 MHz was applied upon acquisition. Due to the small size of the samples, an $Al_2O_3$ hexagonal prism was placed below the sample, and the AE sensor was attached to the prism using high-temperature grease and steel hose clamp. The AE signal transfer was verified using a standardized Hsu-Nielsen test. Synchronization between the deformation and AE datasets was achieved by directly feeding the deformation signal as a parametric input into the AE device during data acquisition.

The AE post-processing entailed AE event individualization following the conventional procedure [15] and was performed using an in-house MATLAB script. The parameters were selected empirically as follows: threshold of 20 µV (note that the background noise was ~12 µV), hit definition time of 500 µs, and hit lockout time of 1000 µs. These values led to an effective detection of a significant number of AE events above the noise level while simultaneously minimizing excessive event overlaps and still capturing plenty of low-amplitude events. The evolution of AE signal median frequency ($f_{med}$) was calculated from the raw AE signal in MATLAB using a window length of 65536 data points (i.e. 32.768 ms), an overlap of 16384 data points (i.e. 8.192 ms), and a Fourier transform length (nfft) of 4096. The resulting $f_{med}$ data were finally smoothed using adjacent averaging of 50 points. Power-law distributions of squared AE event amplitudes ($A^2$, which are roughly proportional to the energy ($E$) of the associated plastic event, $A^2 \sim E$ [16]) were evaluated using a robust maximum-likelihood estimation method implemented in Matlab [17]. Several other AE event individualization parameter sets were also tried within reasonable limits and, in terms of both – the studied AE quantities and PL distributions - led to very similar results, in agreement with the findings of other authors [18].

**Results and Discussion**

The SEM (BSE) micrograph in Fig. 1a shows that the alloy features an indistinct dendritic structure with dendrites (appearing brighter, as highlighted by red arrows). In our previous study [14], it was shown by energy-dispersive X-ray spectroscopy (EDS) that the dendrites are rich in Mo and slightly depleted of Fe. X-ray diffraction (XRD) revealed that the structure is a single-phase bcc solid solution with the lattice parameter $a$=2.997(1) Å, while the scarce black spots likely formed during material processing and were identified by EDS as Al-rich particles (they their low volume fraction rendered them undetectable in the XRD data). The grain structure is relatively homogeneous, the average grain size is ~160 µm, and there is no preferential grain orientation documented by EBSD (Fig. 1b). For detailed microstructural analysis see [14]. In the deformed condition (Fig. 1c), the grain structure remains unchanged, albeit the grains show wavy boundaries and local lattice reorientation due to dislocation activity. There are no deformation twins, as will also be conferred in the following sections.

Fig. 2a shows the deformation curve together with the AE parameters recorded during compression at 300 °C. There are sharp, well-defined stress drops (B-type PLC effect) that correlate with both the AE count rate and AE events amplitude, and, to a good extent, also with the AE median frequency, which decreases with the occurrence of drops, especially during later stages of loading (see also zoomed-in Fig. 2b).

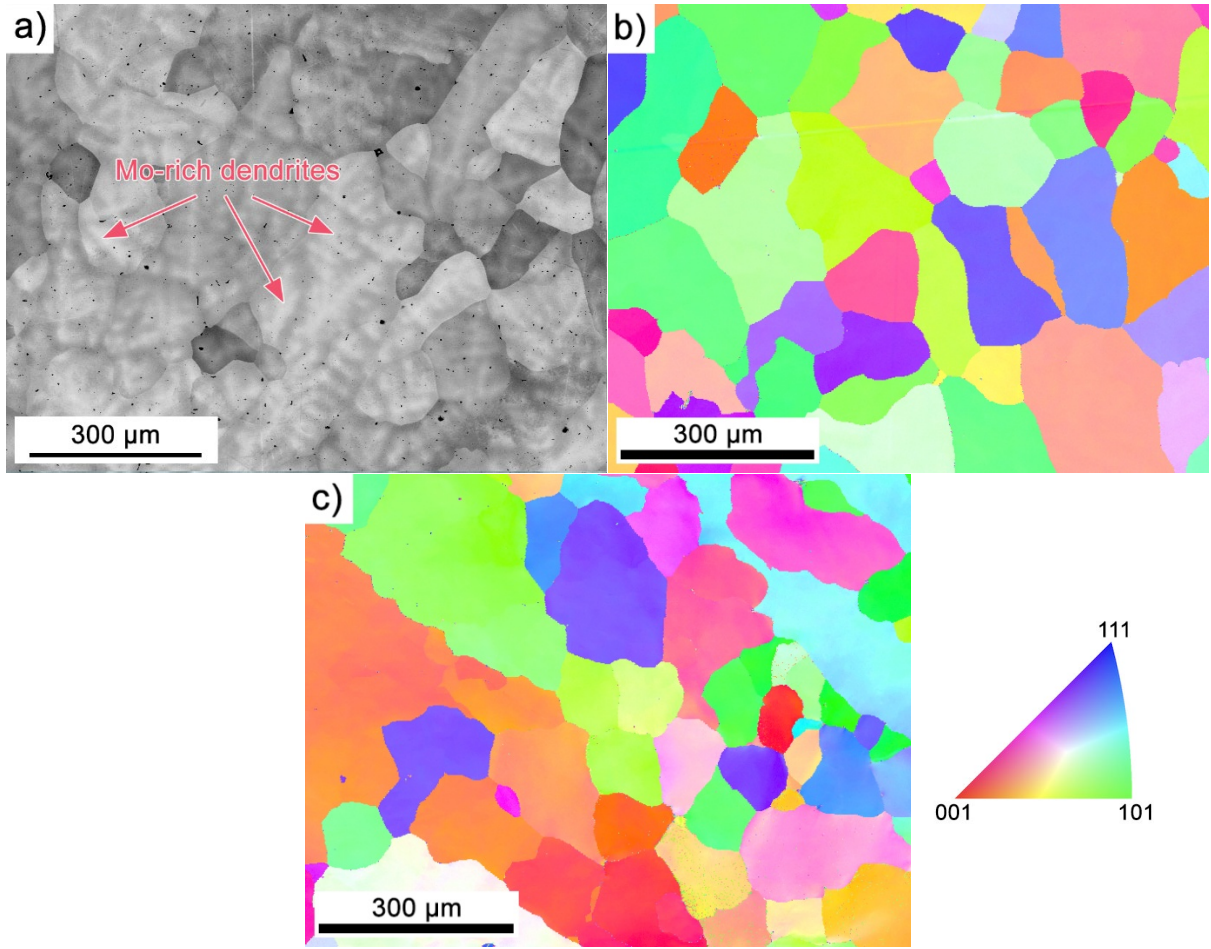

**Fig. 1.** Microstructure of the investigated FeAlCoMo alloy (a) SEM image (BSE signal) before compression, (b) EBSD image before compression, (c) EBSD image after compression at 400 °C (plastic strain of ~8 %).

Fig. 2c represents the same set of quantities recorded during compression at 400 °C. There are some qualitative similarities but also marked differences when compared to deformation at 300 °C. The stress drops feature a duplex character of their magnitudes with well-defined large drops (C-type PLC effect) of ~20 MPa and several (~2-15) small serrations in between them, with a magnitude of ~2-3 MPa. Note that small serrations were shown to also originate in the PLC effect and, interestingly, to be affected by the surface finish quality (polishing homogenizes and increases surface defect density, suppressing the formation of small serrations) [19,20]. The AE parameters strongly correlated with the large drops, but correlations were also observed in the case of small ones, in agreement with [12,20]. Profound correlations, reaching the $f_{med}$ local minima as the large drop appears, take place again only at the end of loading (Fig. 2d). Local decrease in $f_{med}$ was observed in many diluted alloys featuring DSA and reflects the collective unpinning of dislocation and correlated motion of their ensembles (dislocation avalanches) associated with the occurrence of stress drops [21,22].

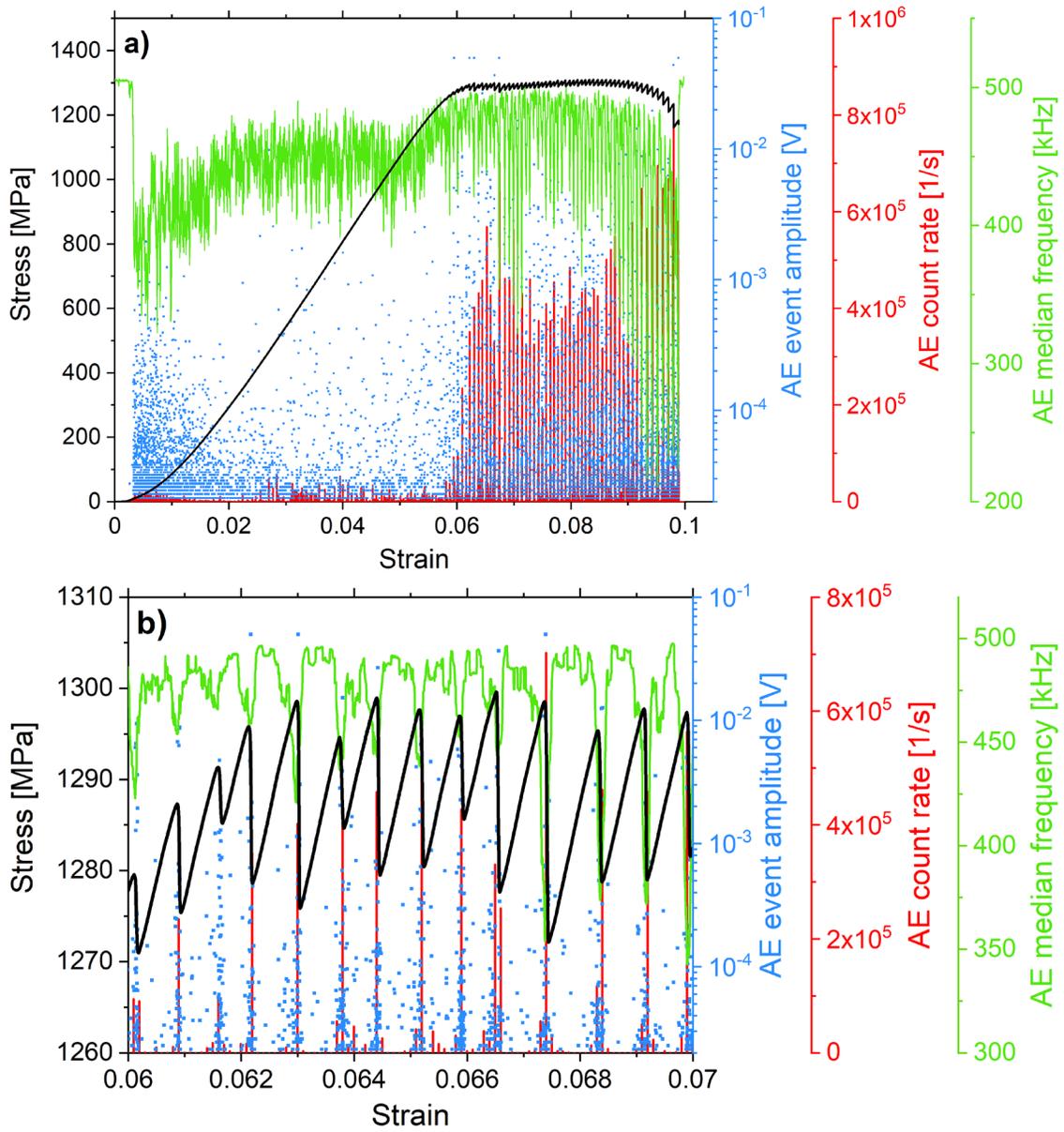

*[Figure continues on the next page.]*

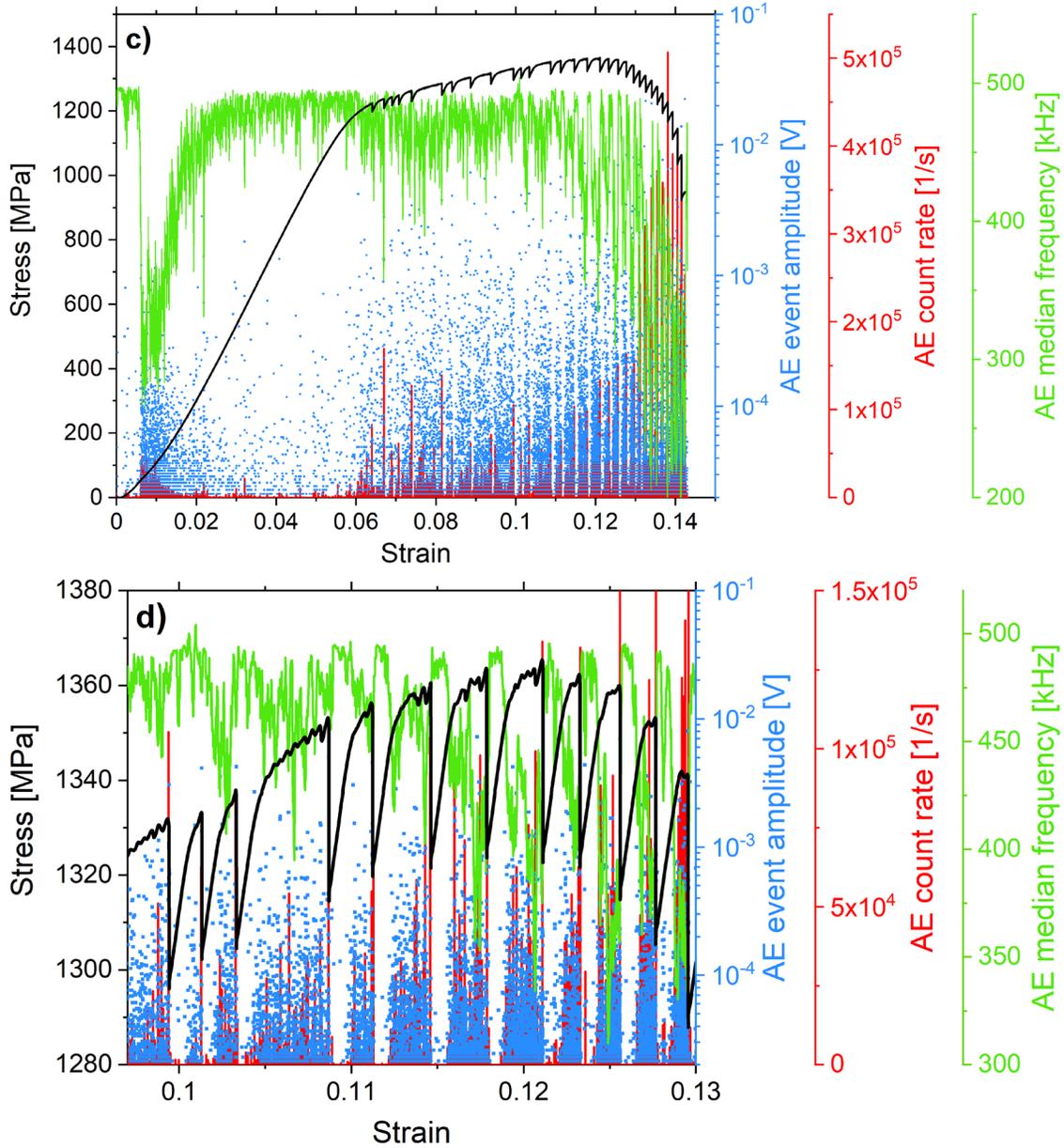

**Fig. 2.** Compression curves of the FeAlCrMo alloy together with the AE parameters recorded during testing at (a) 300 °C, (b) 300 °C zoomed-in, (c) 400 °C, (d) 400 °C zoomed-in.

A salient feature of both tests is the occurrence of AE activity already at the very beginning of loading in the quasi-elastic region (Fig 2ac). There is, thus, no true elastic limit, and irreversible deformation appears already at the beginning of loading [23]. The frequency of the occurrence of these early signals gradually diminishes and they become rather sporadic after approaching the stress of ~500 MPa. To get more insights into this effect, cyclic loading was performed on an additional sample at 400 °C to examine the influence of previous loading cycles on the occurrence of AE. Fig. 3 manifests that AE appears only and exactly when the previous stress level is reached, giving rise to the so-called Kaiser effect [24]. This signifies the unpinning of mobile dislocations upon reaching the stress necessary for their release and their complete exhaustion during initial loading. As observed in the monotonic tests (Fig. 2ac), subsequent AE signals then reappear abruptly during the elastic-plastic transition, i.e. around the yield point, as is common for many metallic materials [24]. The AE signals occur mostly in correlation with load drops (especially in the case of testing at 300 °C), and there is very little AE during reloading, i.e. between the load drops.

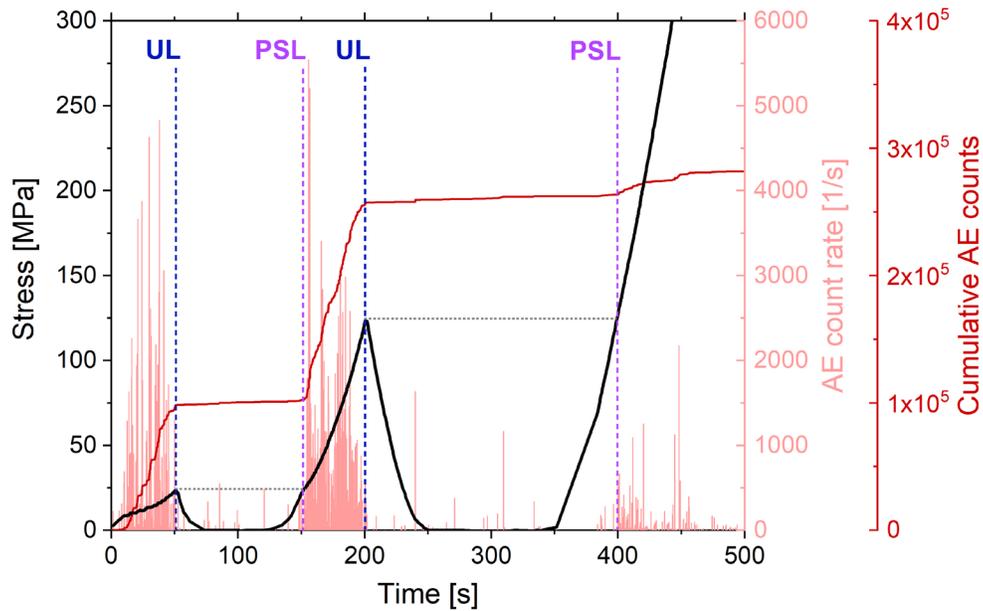

**Fig. 3.** Cycling loading of the FeAlCrMo alloy tested at 400 °C manifesting microplasticity in terms of AE signals appearing already in the elastic stage of loading. Horizontal dotted lines show the stress level reached in the previous cycle. Vertical dashed lines mark the beginning of the unloading (UL) stages and the points where the previous stress level (PSL) is reached again.

It was revealed in the last two decades that most metallic materials show scale invariant (power-law) statistics of the energies $E$ of elementary plastic events (e.g. dislocation avalanches), often probed by AE as an effective high-resolution proxy [16,25–27]. The power-law distribution $P(E) \propto E^{\beta}$ results in the straight line on the double logarithmic plot and is characterized by a single parameter — the exponent $\beta$ (i.e. the line slope). This behavior is obviously strongly non-Gaussian, and no mean physical quantities representing dislocation avalanches can be defined [25]. Furthermore, as mentioned above, $E$ is roughly proportional to the square of the corresponding AE event amplitude, $E \sim A^2$. Fig. 4 shows the statistical distributions of $A^2$ for the samples tested at 300 and 400 °C, and, moreover, distributions of $A^2$ recorded during quasi-elastic loading (i.e. microplasticity) and during plastic loading are depicted separately[1]. It is evident that the power law holds across many orders of magnitude in each instance, with $β$ ~ -1.4 for plastic events at 300 °C, $β$ ~ -1.6 for plastic events at 400 °C, and $β$ ~ -1.7 for microplastic events at both 300 and 400 °C.

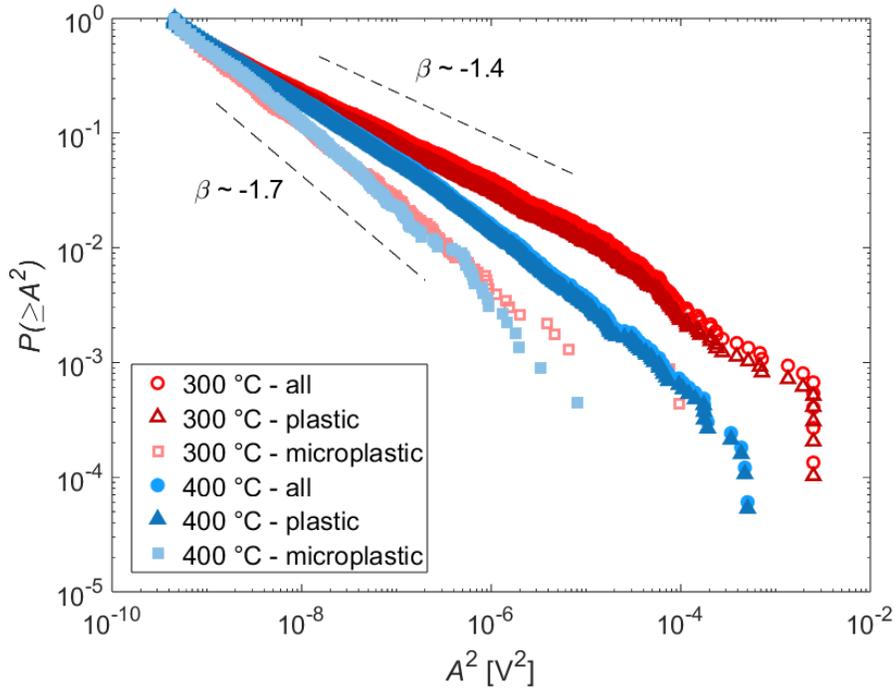

**Fig. 4.** Complementary cumulative density functions (CCDFs) of squared AE event amplitudes, $A^2$, showing power-law distributions for both testing temperatures (300 and 400 °C) and both microplastic and plastic regimes. Note that, graphically, CCDF yields a $β$ value that is one unit higher (i.e. the slope is less steep) than that obtained from the probability density function (PDF). This adjustment has already been accounted for, and the "PDF values" of $β$ are provided to facilitate direct comparison with other studies.

The power-law distributions of $A^2$ have been observed in many pure metals and diluted alloys even without the presence of stress serrations due to DSA, i.e. when the deformation curves appear smooth. The commonly observed exponents have been $β$ ~ -(1.5 – 1.8) for bulk single crystals and micropillars [23,25] and as low as $β$ ~ -(2 – 3) for non-CCA (Al-Mg) alloys exhibiting serrations due to DSA [28]. To the author's best knowledge, the only $β$ reported for CCA ($Al_{0.3}CoCrFeNi$) had the value of -(1.5 – 2.1) depending on the testing conditions [29], which roughly agrees with the observation in this work. On the other hand, there are no reports on $β$ from the occasionally observed microplastic deformation of metallic materials. Higher $β$ observed in DSA-featuring CCAs implies the presence of a greater fraction of bigger events and can be explained in terms of a more complex microstructure of CCAs compared to diluted alloys exhibiting DSA. Even though it is rather challenging to resolve which elements in CCAs act as solutes responsible for DSA, it can be expected that due to high lattice distortion, pinning of dislocations becomes more effective, resulting in a larger magnitude of dislocation avalanches once the critical local stress for their release is achieved. Moreover, it was shown that the presence of Al and even the dendritic structure can play an important role in the DSA of CCAs and, hence, may foster this effect [8,30]. Some authors argue that in CCAs, mechanical twinning can also be related to DSA [8,11]; however, in our study, no twins were identified in the post-mortem micrographs (Fig. 1c).

---

[1] Note that the cut-off in large $A^2$ (i.e. the "big avalanches") is typically related to the finite sample size. Here, especially in the case of the test at 300 °C, the effect is also amplified by the presence of several AE events with amplitudes reaching the upper limit of the voltage range of the AE set-up, cf. Fig 2ab.

Larger $β$ for the test at 300 °C can relate to the dynamics of the observed PLC effect - rapid stress drops and little plastic activity in between them result in the predominance of large events. Slightly lower $β$ during the test at 400 °C stems from the presence of smaller avalanches between the large drops. Different types of the PLC effect (as a function of strain rate or temperature) resulting in different $β$ were also observed elsewhere [28,29]. Finally, $β$ originating in microplastic events is slightly lower ($β$ ~ -1.7), showing a higher fraction of small avalanches and, indeed, very few large ones with $A^2 > 10^{-5}$ $V^2$, unlike during plastic loading (Fig. 4.). Also, when assessing the contribution of microplastic events to the overall distribution, it is energetically relatively insignificant. Interestingly, the distribution of microplastic events during testing at 300 °C and 400 °C is very similar. However, this is not particularly surprising if one realizes that microplastic loading probes the initial microstructure of the material, which is not (much) affected by the test conditions, unlike the "driven" deformation dynamics during plastic loading [23].

Although the power-law characteristics of squared AE amplitudes are often insensitive to specific deformation processes, distinguishing between different dislocation dynamics (e.g., microplastic vs. plastic loading) or their interactions with grain boundaries and other defects remains an intriguing yet highly challenging task. Further studies are needed to elucidate the relationship between evolving dislocation structures and AE signal characteristics, particularly in complex materials such as CCAs, by incorporating complementary in-situ and ex-situ techniques.

## Conclusions

DSA manifesting itself as the PLC effect, being one of the few cases when self-organization of dislocations shows up macroscopically, has been an attractive research topic. It relates to the fundamental understanding of multi-scale characteristics of deformation dynamics, and, from the practical perspective, it is viewed as detrimental to the mechanical performance of the material. We showed by AE analysis that in the investigated material, the characteristics of the PLC effect are similar to those observed in diluted alloys, notwithstanding different degrees of microstructural complexity. The observed scale-invariant behavior (power-law distributions of the energies of plastic events) is a signature of self-organized criticality or synchronization of dislocation processes. Thus, the investigated material is shown to also belong to the predominant class of materials featuring this type of complex dynamic behavior. Moreover, we showed for the first time that scale-invariant behavior also holds for the microplastic regime, i.e. before the macroscopic yielding. The scarcity of literature data, however, calls for further research endeavors and extension of the current understanding of deformation dynamics to the rich arena of CCAs with complex microstructures.

## Acknowledgements

This work received funding from the Czech Science Foundation (project No. 21-03194S). T.T. would like to acknowledge financial support from the Charles University Grant Agency (project No. 354922). Financial support by OP Johannes Amos Comenius of the MEYS of the Czech Republic (project No. CZ.02.01.01/00/22_008/0004591) is also gratefully acknowledged.

## Data availability

The data that support the findings of this study are openly available in the Zenodo repository at https://doi.org/10.XXXX/zenodo.YYYYYYYY [the link will be updated later] under the CC-BY 4.0 license.

## License